# Calculation of DOS-Dependent Channel Potentials in FETs in the Saturation Condition

Woo-young So[1], David V. Lang[1], and Arthur P. Ramirez[1, 2]

[1]*Columbia University, New York, NY 10027*
[2]*Bell Laboratories, Lucent Technologies, 600 Mountain Avenue, Murray Hill, NJ 07974*

ABSTRACT

We calculated functions for the potential, $V_{ch}$, in the channel of a field effect transistor with various densities of the localized states (DOS) by using a device simulator. In the saturation condition, $V_{ch}$ is found to fit satisfactorily to the analytical function, $V_{ch}(y)=V_g(1-(1-y/L)^k)$, where $V_g$ is the gate bias, $y$ is the position along the channel, and $L$ is the channel length. The power $k$, which depends on the DOS, ranges from 0 to 0.5 even with an extensive variation of DOS. As a result, the total induced charge density in the saturation condition can be expressed a way similar to the linear condition by introducing a correction factor, $\alpha$, ranging between 0.65 and 1.0.

Since Field-Effect-Transistors (FETs) were invented in the late 1940s, the devices have become ubiquitous. Metal-Insulator-Semiconductor (MIS) FETs are especially widely used, and the theory of device operation is mature. Since most electrical properties and device characteristics of MISFETs can be expressed through the electrical potential in the channel, such potential functions have been extensively modeled. Generally, turn-on states of MISFETs can be divided into two regions depending on the bias conditions of the terminals: the linear and the saturation conditions; whereas the linear condition can be described by a relatively simple potential function, the saturation condition leads to more complicated solutions. Though Pao and Sah have already derived the channel potential in the saturation condition,[1] this treatment is inapplicable to materials with high densities of localized states (DOS). Localized states will strongly influence device performance, as discussed in the accompanying letter.[2] Specifically, the immobile charge in the active channel alters the total charge which, in turn, determines the potential distribution along the channel. It is the purpose of this note to quantitatively model the channel potential at saturation for arbitrary DOS.

When the FET is operated in the saturation condition, the channel potential, $V_{ch}$, decreases slowly across the channel and reaches $V_{ch}=V_g$ at a position close to the drain. The remaining potential, $|V_d|-|V_g|$, drops abruptly near the drain only within a very narrow region, called the pinch-off region. Assuming the pinch-off region is negligible compared to the channel length, the boundary conditions of $V_{ch}$ become $V_{ch}(y=0) = 0$ and $V_{ch}(y=L) = V_g$, where y is the longitudinal coordinate along the channel from the source to the drain. Thus, a solution to Poisson's equation for $V_{ch}$ with respect to y, which satisfies these boundary conditions, can be expressed by the analytical form

$$V_{ch}(y) = V_g(1-(1-y/L)^k), \tag{1}$$

where *L* is the channel length and *k* is a constant. Here, only *k* is dependent on the trapped charges since the other parameters are given by the device structure and bias conditions. As we show below, this function gives a relatively good fit to the results of a full 2D device simulation.

In order to solve Poisson's equation in the presence of the localized states with charges, we use a 2D device simulator, Silvaco ATLAS.[a] For the device simulation, Si is used as the semiconducting material with the effective DOS in the valence band and conduction band changed, $N_v = N_c = 1.08 \times 10^{20}$ $cm^{-3}$, which is equivalent to our two-dimensional DOS in the organic semiconductors with a depth of 1*nm*. In addition, a 700nm SiO$_2$ layer (with built-in materials parameters) is used as the gate dielectric to emulate the 500nm parylene film typically used in this study. In the device simulator, the "TFT" module allows us to incorporate an exponential energy distribution function of localized states in the bandgap of Si: *NTD·exp(-E/WTD)*, where *NTD* and 1/*WTD* correspond to $N_{1,0}$ or $N_{2,0}$ and $\beta_1$ or $\beta_2$ in our letter, respectively. The total volume charge density induced by the gate bias is then given by

$$\rho_{total}(V_g) = \rho_{trapped} + \rho_{free} = NTD \cdot WTD \cdot \exp(-F/WTD) + N_v \cdot \exp(-F/kT). \quad (2)$$

Since $\rho_{trapped}$ has the same analytical expression as $\rho_{free}$ when *WTD* is equal to *kT*, a reference value of the pre-exponential factor, $N_{gap0}$, in two-dimensional states is used with the 1nm monolayer scale, such that $\rho_{trapped}$ and $\rho_{free}$ are the same at 300K with gate biases used in the experiment. Thus, the two-dimensional gap states in the simulation can be considered to be $N_{gap0} \cdot exp(-\beta \cdot E)$. After solving Poisson's equation with varying $N_{gap0}$ and $\beta$ values, $V_{ch}(y)$ is plotted in Fig. 1 at 1*nm* depth perpendicular to the interface between the semiconductor and dielectric material.

---

[a] http://www.silvaco.com/products/device_simulation/atlas.html

Fig. 1 shows the numerical solution of $V_{ch}(y)$ with $N_{gap0}=1.0\times10^{15} cm^{-2}eV^{-1}$ and $\beta=28eV^{-1}$. As explained before, the channel and pinch-off region are considered separate regions, and the narrow pinch-off region can be estimated less than 3% of the whole device length at $V_g$=-10V. Hence, the boundary conditions of $V_{ch}$ can be validated, and the power, k, can be found by fitting the numerical solution to Eq. 1. As presented in Fig. 1, $V_{ch}$ is sufficiently expressed in the analytical form with k=0.389 regardless of $V_g$. Once $V_{ch}$ is fitted by the analytical function, the total induced charge, $n_{total}$ is given by

$$n_{total} = C/q \int_0^L V_g(1-y/L)^k \, dy/L = -CV_g/(q \cdot (k+1)), \qquad (3)$$

which means that the fraction of total induced charge density in the linear condition[b] is equal to $1/(k+1)$, and it can be defined as a correction factor, $\alpha$.

As shown in Fig. 2, the correction factor $\alpha$ is calculated for wide variety of DOS values. Here, $\alpha$ ranges from 0.65 to 1 with various parameters of $N_{gap0}$ and $\beta$. As the DOS decreases due to a lower $N_{gap0}$ and higher $\beta$, $\alpha$ approaches 0.65, whereas α increases up to the limiting value of 1 for higher DOS. At the lower DOS limit around $N_{gap0}=1.0\times 10^{11}cm^{-2}eV^{-1}$, $\alpha$ becomes effectively constant at 0.65. $N_0=1.0\times 10^{16}cm^{-2}eV^{-1}$ is comparable to the molecular surface density with the thermal activation energy of an oligomer semiconductor: the molecular surface density of pentacene with the 1nm monolayer is $2.9\times 10^{14} cm^{-2}$.

In summary, we calculate the channel potential in the saturation condition under a variety of DOS conditions. The channel potential can be fit to the simple analytical function by introducing the power, k, in the function, which is dependent on the DOS. The developed function may provide a useful model in device or SPICE modeling. In addition, the parameter bridges the total induced charge densities in the linear and the saturation conditions.

---

[b] $N_{total,linear} = -CV_g/q$

We acknowledge the technical assistance of Silvaco Korea Co. Ltd. through the device simulator.

Figure Captions

Figure 1. Comparison between device-simulated numerical solution and a fitting function of channel potential. A 150μm-long channel, typical of our devices, is considered under $V_d$=-40$V$ used in this work. The coordinates 0 and 150 correspond to the source and drain position respectively.

Figure 2. Correction factor for total induced charge in the saturation condition.

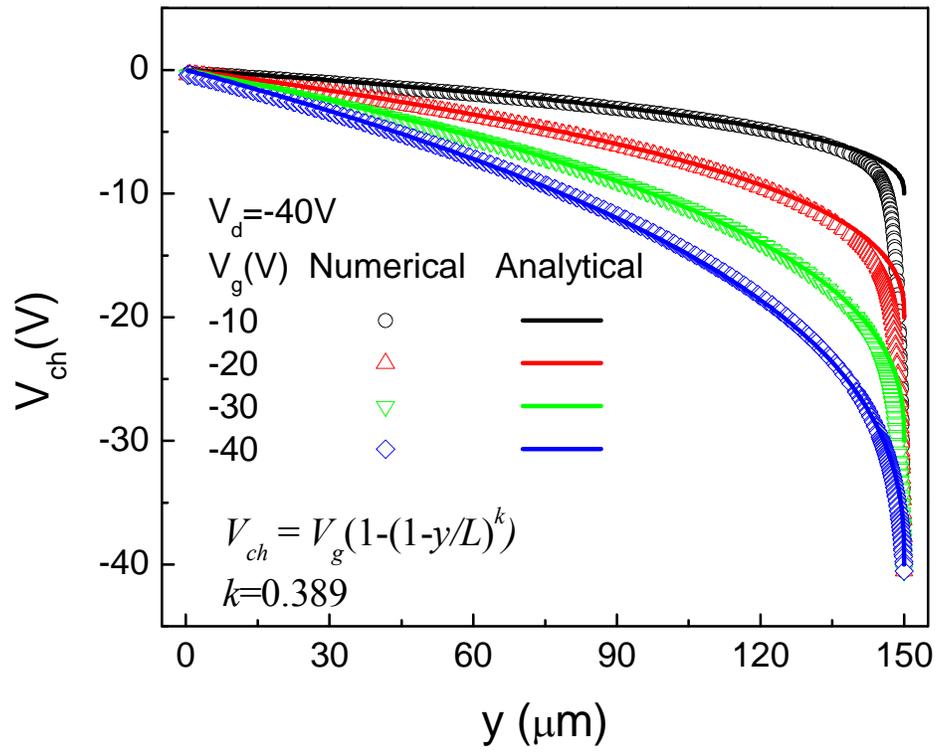

**Fig. 1**

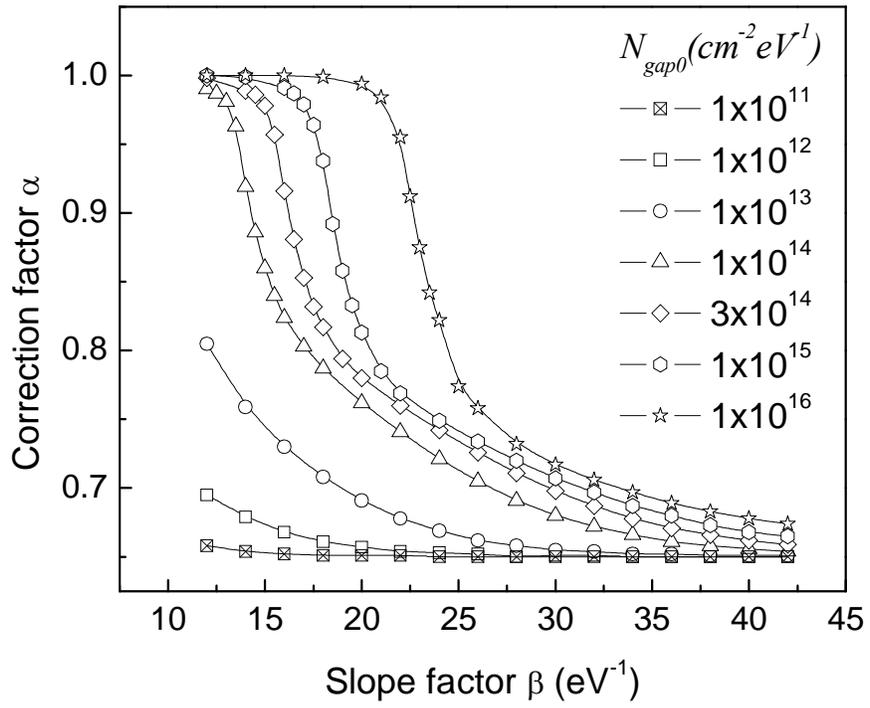

**Fig. 2**